# Low-Resistance 2D/2D Ohmic Contacts: A Universal Approach to High-Performance WSe$_2$, MoS$_2$, and MoSe$_2$ Transistors

Hsun-Jen Chuang[1], Bhim Chamlagain[1], Michael Koehler[2], Meeghage Madusanka Perera[1], Jiaqiang Yan[2,3], David Mandrus[2,3], David Tománek[4], and Zhixian Zhou[1,*]

[1] Physics and Astronomy Department, Wayne State University, Michigan 48201, USA
[2] Department of Materials Science and Engineering, The University of Tennessee, Knoxville, TN 37996
[3] Materials Science and Technology Division, Oak Ridge National Laboratory, Oak Ridge, TN 37831
[4] Physics and Astronomy Department, Michigan State University, East Lansing, Michigan 48824, USA

[*] Email of the corresponding author: zxzhou@wayne.edu

Abstract

We report a new strategy for fabricating 2D/2D low-resistance ohmic contacts for a variety of transition metal dichalcogenides (TMDs) using van der Waals assembly of substitutionally doped TMDs as drain/source contacts and TMDs with no intentional doping as channel materials. We demonstrate that few-layer WSe$_2$ field-effect transistors (FETs) with 2D/2D contacts exhibit low contact resistances of ~ 0.3 kΩ μm, high on/off ratios up to > 10$^9$, and high drive currents exceeding 320 μA μm$^{-1}$. These favorable characteristics are combined with a two-terminal field-effect hole mobility μ$_{FE}$ ≈ 2×10$^2$ cm$^2$ V$^{-1}$ s$^{-1}$ at room temperature, which increases to >2×10$^3$ cm$^2$ V$^{-1}$ s$^{-1}$ at cryogenic temperatures. We observe a similar performance also in MoS$_2$ and MoSe$_2$ FETs with 2D/2D drain and source contacts. The 2D/2D low-resistance ohmic contacts presented here represent a new device paradigm that overcomes a significant bottleneck in the performance of TMDs and a wide variety of other 2D materials as the channel materials in post-silicon electronics.

KEYWORDS: MoS$_2$, WSe$_2$, MoSe$_2$, field-effect transistor, two-dimensional, ohmic contact

The layered nature of transition metal dichalcogenides (TMDs) allows for easy cleavage and formation of ultrathin layers, which are being considered as suitable semiconducting counterparts to semi-metallic graphene and may lead to flexible electronics and optoelectronics applications.[1-5] However, fabrication of high-performance transistors of TMDs including $WSe_2$, $MoS_2$, and $MoSe_2$ has been a major challenge in 2D electronics.[6, 7] The performance of current metal-contacted TMDs is limited by the presence of a significant Schottky barrier (SB) in most cases.[8-11,12] In silicon-based electronics, low-resistance ohmic contacts are achieved by selective ion implantation of drain/source regions below metal electrodes. In this way, the contact barrier width between the metal electrodes and degenerately doped source and drain regions is significantly reduced. Unfortunately, the ultrathin body of monolayer and few-layer TMDs prohibits effective doping by ion implantation. Various other doping methods such as surface charge transfer doping[8, 13-15] and substitutional doping[16, 17] have also been developed by different groups during the past few years to reduce the Schottky barrier width and thus reduce the contact resistance of TMD devices. However, most of these doping methods suffer from poor air or thermal or long-term stability. In this respect, substitutional doping appears to offer a suitable alternative, since dopants secured by covalent bonding (e.g. Nb doped $MoS_2$) during the material synthesis yield devices with excellent air and thermal stability.[17] However, the limitation of conventional substitutional doping during synthesis is the inability to form a spatially abrupt doping profile, which defines the drain, the channel and the source regions, and which is needed for low power, high-performance electronics.

To date, various innovative strategies to reduce the contact resistance such as use of graphene contacts[3, 18-20] and phase-engineering,[21, 22] are still deficient as they do not offer true ohmic contact behavior or have insufficient thermal stability. Nearly barrier-free contacts to $MoS_2$ have been achieved by using graphene as contact electrodes because the Fermi level of graphene can be effectively tuned by a gate voltage to align with the conduction band minimum ( CBM ) of $MoS_2$, which minimizes the Schottky barrier height (SBH). [23, 24, 25] Still, a significant SBH is usually formed between graphene and $WSe_2$ because the work function of graphene is close to the middle of the band gap in $WSe_2$.[18, 19] We have previously used the extremely large electric double layer (EDL) capacitance of an ionic liquid (IL) gate to minimize the SBH by tuning the work function of graphene at the graphene/$WSe_2$ interface within an extremely large range.[18] As a result, we have formed for the first time, in a single device structure, $WSe_2$-based FETs of both *n*- and *p*-type that display low-resistance contacts (down to ~ 2 kΩ.µm) and a high carrier mobility (> 300 $cm^2V^{-1}s^{-1}$ at 77 K). However, for realistic device applications, methods to achieve more permanent, air-stable and thermally stable ohmic contacts with an order of magnitude lower contact-resistance are needed. Significant SBH reduction can also be achieved by locally inducing the metallic 1T phase $MoS_2$ on semiconducting 2H phase $MoS_2$ flakes, which can be attributed to the atomically sharp interface between the 1T and 2H phases and to the fact that the work function of the 1T phase is very close to the CBM of the 2H phase. [21] However, a finite SBH is expected to arise for the electron (hole) channel, when the work function of the 1T metallic phase does not line up with the CBM (Valence band maximum; VBM) of the 2H phase. For instance, a large SBH is expected for the hole channel of $MoS_2$ FETs using this phase-engineering contact strategy because of the large offset between the work function of the 1T phase and the VBM of the 2H phase. Furthermore, the 1T phase MoS$_2$ is thermally unstable above 100 $^o$C. The availability of a variety of semiconducting TMDs such as MoSe$_2$, WS$_2$ and WSe$_2$ with different band structures and charge neutrality levels offers additional distinct properties and opportunities for device applications. [5, 6, 8, 9, 13, 26-37] However, the variation of electron affinity, band gap, and band alignments also presents significant challenges to contact engineering. To unlock the full potential of TMDs as channel materials for high-performance thin-film transistors, highly effective and versatile contact strategies for making low-resistance ohmic contacts are needed.

In this letter we present a new strategy that utilizes 2D/2D vertical junctions to engineer low-resistance ohmic contacts, which turn TMDs including WSe$_2$, MoS$_2$ and MoSe$_2$ into high-performance transistors. 2D/2D junctions with atomically sharp interfaces can be created by van der Waals assembly of 2D crystals without the constraints of atomic commensurability.[38, 39] We demonstrate that 2D/2D contacted FETs consisting of an undoped few-layer WSe$_2$ channel and degenerately $p$-doped WSe$_2$ drain and source contacts exhibit low contact resistances of ~ 0.3 kΩ μm, high on/off ratios up to > 10$^9$, and high drive currents exceeding 320 μA μm$^{-1}$. Furthermore, low resistance ohmic contacts achieved in our devices enable the investigation of intrinsic channel properties of TMD materials. Our WSe$_2$ devices with 2D/2D contacts display a two-terminal field-effect hole mobility μ$_{FE}$ ≈ 2.2×10$^2$ cm$^2$ V$^{-1}$ s$^{-1}$ at room temperature, which increases to about 2.1×10$^3$ cm$^2$ V$^{-1}$ s$^{-1}$ at 5 K. Similarly, record high two-terminal field-effect hole mobility up to 2.8×10$^3$ cm$^2$ V$^{-1}$ s$^{-1}$ (at cryogenic temperatures) has been observed in MoS$_2$ and MoSe$_2$ FETs with degenerately $p$-doped MoS$_2$ contacts formed by van der Waals assembly.

Figure 1a,b presents a schematic diagram and optical micrograph of a WSe$_2$ FET device

composed of degenerately *p*-doped WSe$_2$ (Nb$_{0.005}$W$_{0.995}$Se$_2$) 2D drain/source electrodes in contact with a 2D WSe$_2$ channel with no intentional doping. Devices containing TMDs such as WSe$_2$ were fabricated by artificially stacking mechanically exfoliated flakes of degenerately *p*-doped TMDs, considered as electrodes, on top of an undoped TMD channel material using a dry transfer method.[18] Subsequently, metal electrodes, consisting of 5 nm Ti / 50 nm Au, were formed by deposition on top of the degenerately doped TMD contacts (see the Methods and Sections 1 and 2 of the Supplementary Information). To preserve its intrinsic electrical properties, the TMD channel material was encapsulated in hexagonal boron nitride (hBN).[18, 40] Similar to degenerately doped silicon in Si electronics, ohmic contacts with low contact resistance < 0.2 kΩ µm is also achievable between degenerately doped TMDs and the top metal electrodes (see Section 3 of the Supplementary Information). Consequently, the total contact resistance critically depends on the resistance of the 2D/2D junctions between the degenerately doped TMDs, acting as source/drain electrodes, and the undoped TMD channel material.

The band diagram and working principle of the 2D/2D contacts are illustrated in Fig. 1c. The difference in work function between the undoped channel and the degenerately doped drain/source, caused by the different carrier densities, creates a band offset across the 2D/2D interface. In conventional 3D semiconductor junctions, the band offset is usually well-defined by the covalent bonds at the junction interface. Since the inter-layer interaction in 2D TMDs and their junctions is much weaker, the band offset can be electrostatically tuned by a back-gate voltage.[38, 39] We take advantage of this unique property of 2D/2D junctions to form spatially sharp, tunable, true ohmic contacts to TMDs.

As seen in the top panel of Fig. 1c, there are no free carriers in the channel in the off-state at the back-gate voltage $V_{bg}$=0 V. Increasing the negative back-gate voltage shifts all bands in the channel material up, whereas the bands in the degenerately doped electrodes are unaffected. The modified band alignment introduces holes in the channel material, as illustrated in the bottom panel of Fig. 1c. In the on-state, achieved at gate voltages exceeding the threshold ($|V_{bg}|>|V_{th}|$), the contact barrier at the interface essentially vanishes, leading to a low-resistance contact.

Figure 1d,e shows the room-temperature transfer and output characteristics of a 5-layer WSe$_2$ FET that is encapsulated in hBN. This ~3.5 nm thick device is contacted by degenerately $p$-doped WSe$_2$ (Nb$_{0.005}$W$_{0.995}$Se$_2$) and measured using a Si back gate. The gate dielectric consists of 40 nm thick hBN on 280 nm thick SiO$_2$. It shows clear $p$-type behavior with an exceptionally high on/off ratio exceeding $10^9$ at $V_{ds}$ = -1 V and a subthreshold swing of ~460 mV/dec, which can be further reduced to the near-ideal value of ~63 mV/dec by using a top gate with hBN gate dielectric (see Section 5 of the Supplementary Information). The gate voltage range can also be significantly reduced by using thinner and high-κ dielectrics. The high on/off ratio can be partially attributed to the significant enhancement of the on-current that is enabled by the low-resistance 2D/2D contacts. As shown in Fig. 1e, the on-state drain current is linear at all back-gate voltages, indicating ohmic behavior. Although we present results on $p$-type TMD transistors in this work, we have also achieved $n$-type behavior using heavily $n$-doped TMDs as drain and source contacts (see Section 6 of the Supplementary Information). This is an important advantage of the proposed 2D/2D contact strategy because availability of both $p$-type and $n$-type 2D transistors with low-resistance

ohmic contacts is crucial for CMOS applications.

We quantify the contact resistances of the 2D/2D contacts using the transfer length method (TLM). Figure 2a,b shows the schematic diagram and an optical micrograph of a WSe$_2$ test structure for TLM measurements, consisting of an ~7 nm thick undoped WSe$_2$ channel, outlined by the dash-dotted lines, connected to ~21 nm thick $p$-doped WSe$_2$ (Nb$_{0.005}$W$_{0.995}$Se$_2$) drain and source contacts with varying gap spaces, outlined by the dashed lines. From the y intercept of the linear fit to the total resistance as a function of the channel length, we extract a contact resistance of ~ 0.3 kΩ µm, as shown in Fig. 2c. Consistent results are obtained in WSe$_2$ devices with channels thinner than 10 nm (see Section 4 of the Supplementary Information). The contact resistance $R_C$ here is significantly lower than what is found for graphene/WSe$_2$ contacts (~2 kΩ µm)[18] and compares favorably with the best results achieved on TMD devices (~ 0.2 - 0.7 kΩ µm).[16, 21, 41]

Figure 2d shows the output characteristics of a ~ 7.0 nm thick short-channel WSe$_2$ device with $L \approx 0.27$ µm, $W \approx 0.50$ µm and 2D/2D contacts, which is part of the test structure for TLM measurements, shown in Fig. 2b. The device exhibits large drive currents exceeding 320 µA µm$^{-1}$, which are comparable to the highest drive currents achieved in few-layer TMD devices.[16] It is worth noting that even at the large values $V_{bg}$ = -130 V and $V_{ds}$ = -1.5 V, $I_{ds}$ has not yet reached saturation, which indicates that still higher drive currents should be achievable.

Low-resistance 2D/2D contacts also enable us to investigate the intrinsic properties of the channel. Figure 3a presents the temperature-dependent two-terminal conductivity of another WSe$_2$ device that is 3.5 nm thick, ≈14.8 µm long, ≈4.7 µm wide, and passivated by h-BN. The two-terminal conductivity is defined by $\sigma=I_{ds}/V_{ds}\times L/W$, where $L$ is the length and $W$ the width of the channel. With increasing hole concentration, the WSe$_2$ device displays a crossover from an insulating regime, where the conductivity increases with increasing temperature, to a metallic regime, where the conductivity decreases with increasing temperature. This metal-insulator-transition (MIT) can be more clearly seen in the corresponding temperature-dependent conductivity, displayed in the inset of Fig. 3a, and occurs at a critical conductivity of $\sim e^2/h$, consistent with the MIT observed previously in MoS$_2$, MoSe$_2$ and WSe$_2$.[18, 34, 42] The MIT observed here is unlikely a hysteretic effect because the transfer curves measured with opposite gate-sweeping directions overlap to a large degree (see also Section 7 of the Supplementary Information). As seen in the inset of Fig. 3c, the output characteristics of the device remain linear down to 5 K, confirming a true ohmic contact at the 2D/2D interface that is free of a Schottky barrier. Consequently, the on-state conductivity at $V_{bg}$ = -80 V increases monotonically by a factor of ~6 as the temperature decreases from 300 K to 5 K. The temperature-dependent field-effect hole mobility of the WSe$_2$ device is shown in Fig. 3c. The values have been extracted from the linear region of the conductivity curves in the metallic state at -80 V < $V_{bg}$ < -50 V using the expression $\mu_{FE}=(1/C_{bg})\times(d\sigma/dV_{bg})$, where $C_{bg}$ is the geometric back-gate capacitance of 27 nm thick hBN on 285 nm thick SiO$_2$ based on the parallel plate capacitor model. This geometric capacitance is consistent with the back-gate capacitance of a similarly hBN-encapsulated WSe$_2$ Hall bar device determined by Hall measurement (see Section 8 of the Supplementary Information). As the temperature decreases from room temperature to 5 K, the hole mobility for the WSe$_2$ device increases from $\sim 2.2\times 10^2$ cm$^2$V$^{-1}$s$^{-1}$ to about $2.1\times 10^3$ cm$^2$V$^{-1}$s$^{-1}$. This mobility increase with decreasing temperature, along with the large mobility values, suggests strongly that the hole transport in the device is limited by phonons in the channel.

Next, we demonstrate that the 2D/2D contact strategy can also be used to achieve low-resistance contacts for the hole channel of MoS$_2$ FET devices, which has been a major challenge since hole injection across the metal/MoS$_2$ interface has been obstructed by a large Schottky barrier.[35, 43, 44] Figure 3b shows the two-terminal conductivity of an MoS$_2$ FET device consisting of a 6.8 nm thick MoS$_2$ channel with no intentional doping, contacted by degenerately $p$-doped MoS$_2$ (Nb$_{0.005}$Mo$_{0.995}$S$_2$) drain and source electrodes. In contrast to MoS$_2$ devices with conventional metal contacts, which overwhelmingly display $n$-type behavior, the above MoS$_2$ device exhibits $p$-type behavior. The temperature-dependent conductivity of the $p$-type MoS$_2$ device also shows an MIT, as seen in the inset of Fig. 3b. We observe an ~13 fold increase of the on-state conductivity as the temperature decreases from 300 K to 5 K and linear output characteristics down to 5 K, depicted in the inset of Fig. 3d, indicating a barrier-free contact with a low contact resistance. The larger negative threshold voltage observed in our $p$-type MoS$_2$ than in $p$-type WSe$_2$ FETs can be attributed to the higher level of unintentional $n$-doping in the MoS$_2$ channel material. As seen in Fig. 3d, the field-effect hole mobility of the MoS$_2$ device increases from ~$1.8 \times 10^2$ cm$^2$V$^{-1}$s$^{-1}$ to about $2.8 \times 10^3$ cm$^2$V$^{-1}$s$^{-1}$ as the temperature decreases from room temperature to 5 K. To the best of our knowledge, the low-temperature mobility values observed in our $p$-type WSe$_2$ and MoS$_2$ devices represent record high two-terminal hole mobility values in few-layer TMDs,[8, 13, 31] which can be attributed to both low-resistance 2D/2D ohmic contacts and hBN channel passivation.

To further demonstrate the versatility of the new contact paradigm and to elucidate the nature of 2D/2D contacts, we also investigated TMD devices with 2D/2D hetero-contacts, where the drain/source electrodes and the channel consist of different TMD materials. Figure 4a,b presents the output characteristics of representative devices consisting of an $MoSe_2$ channel and degenerately $p$-doped $MoS_2$ ($Nb_{0.005}Mo_{0.995}S_2$) and $WSe_2$ ($Nb_{0.005}W_{0.995}Se_2$) drain/source contacts, measured at 80 K. Whereas the $MoSe_2$ device in Fig. 4a with $p$-doped $MoS_2$ contacts shows linear output characteristics indicative of ohmic behavior, the $MoSe_2$ device in Fig. 4b with $p$-doped $WSe_2$ contacts displays a strongly nonlinear behavior. The nonlinearity and reduction of $I_{ds}$ suggest the presence of a significant contact barrier in this case. Figure 4c,d presents two-terminal conductivity as a function of gate voltage for the same devices at different temperatures. The $MoSe_2$ device with $p$-doped $MoS_2$ contacts exhibits a similar behavior as the $WSe_2$ and $MoS_2$ devices presented in Fig. 3. This similarity includes the presence of an MIT, an increase in on-state conductivity, measured at $V_{bg}$ = -100 V and $V_{ds}$ = -10 mV, by a factor of ~19 as the temperature decreases from 300 K to 5 K. As seen in the inset of Fig. 4c, the phonon-limited two-terminal mobility increases from ~$1.0 \times 10^2$ $cm^2V^{-1}s^{-1}$ at 300 K to about $2.4 \times 10^3$ $cm^2V^{-1}s^{-1}$ at 5 K. In sharp contrast, the two-terminal conductivity in the $MoSe_2$ device with $p$-doped $WSe_2$ drain/source contacts decreases rapidly as the temperature decreases from 240 K to 80 K, indicating contact-limited charge transport.

The drastically different behavior observed in devices with 2D/2D hetero-contacts can be attributed to the differences in band alignments between the channel and contact materials.[45] In the $MoSe_2$ directly underneath the drain/source contacts ( region I in Fig. 4e,f), the VBM of the MoSe$_2$ channel is aligned to the VBM of the contact material (*p*-doped MoS$_2$ or *p*-doped WSe$_2$) by gate voltage when the channel is turned on, leading to a vanishing contact barrier at the 2D/2D vertical contact. However, this band alignment along the vertical direction in region I creates a band offset along the lateral direction parallel to the channel.[46, 47] Because the Fermi level in the entire system must be the same in equilibrium, a local up-turn occurs in the valence band in the lateral interface region, called region II, in the case the VBM of the isolated MoSe$_2$ channel is above the VMB of the isolated *p*-doped MoS$_2$ drain/source contacts. This is accompanied by a flow of holes towards the lateral interface region II, where the hole accumulation builds up a local electric filed that eventually prevents further charge redistribution. Since hole accumulation in region II does not hinder hole transport, ohmic behavior is observed. In the case the VBM of the MoSe$_2$ channel is below the VBM of *p*-doped WSe$_2$ drain/source contacts, a downward band bending occurs in the lateral interface region II, which is accompanied by a flow of holes away from this lateral interface region. The hole depletion in region II acts as a barrier hindering hole transport, leading to non-ohmic behavior. We obtained consistent results in multiple *p*-type TMD FETs with 2D/2D hetero-contacts: ohmic contacts are formed when the VBM of the channel material is above the VBM of the contact material and non-ohmic behavior occurs when the VBM of the channel material is below that of the contact material. (see Sections 10 and 11 of the Supplementary Information).

In summary, we have developed a novel 2D/2D contact strategy to achieve high-quality ohmic contacts for MoS$_2$, MoSe$_2$ and WSe$_2$ FETs. The low-resistance ohmic contacts lead to drastically improved device performance, including on/off ratios up to >10$^9$, drive currents >320 µA µm$^{-1}$, and two-terminal extrinsic field-effect mobilities up to 2.8×10$^3$ cm$^2$V$^{-1}$s$^{-1}$ at cryogenic temperatures. The newly developed contact engineering approach is applicable to a wide range of 2D materials for both *p*-type and *n*-type transistors, compatible with conventional semiconductor processes, and may be implemented in roll-by-roll production of flexible electronics with the development of large scale synthesis techniques.

**Methods**

All crystals of degenerately doped and undoped TMDs used in this work were synthesized by chemical vapor transport except for undoped MoS$_2$ crystals, which were purchased from SPI Supplies. Optical microscopy and Park-Systems XE-70 noncontact mode atomic microscopy (AFM) were used to identify and characterize thin TMD flakes.

To Fabricate TMD devices with 2D/2D contacts, thin flakes of degenerately doped and undoped ultrathin TMDs were mechanically exfoliated from bulk crystals. The degenerately doped TMD flakes, forming the drain and source electrodes, were then artificially stacked using a dry transfer method on top of undoped TMDs flakes, which form the channel. Metal electrodes were then fabricated on top of the degenerately doped source and drain contact regions by standard electron beam lithography and subsequent deposition of 5 nm of Ti and 50 nm of Au (see also Section 2 of the Supplementary Information).

Electrical properties of the devices were measured by a Keithley 4200 semiconductor parameter analyzer in a Lakeshore Cryogenic probe station under high vacuum (1×10$^{-6}$ Torr) or in a Quantum Design PPMS.

**Supporting Information**

Supporting information contains details of the fabrication process for TMD FETs with 2D/2D contacts, additional transport data on TMD devices with 2D/2D contacts, contact resistance between metal and degenerately *p*-doped $WSe_2$, detailed working principle of 2D/2D hetero-contacts, and long term air stability of TMD devices with 2D/2D contacts.

**Acknowledgements**

H.C., B.C., M.M.P, and Z.Z. acknowledge partial support by NSF grant number DMR-1308436 and the WSU Presidential Research Enhancement Award. D.T. acknowledges partial support by the NSF/AFOSR EFRI 2-DARE grant number #EFMA-1433459. M.K. and D.M. acknowledge support from the Gordon and Betty Moore Foundation's EPiQS Initiative through Grant GBMF4416. JY acknowledges support from the National Science Foundation through award DMR-1410428.

*Conflicts of interest:* The authors declare no competing financial interest.

**Figure Captions**

**Figure 1 | Design and characteristics of a high-performance WSe$_2$ FET with 2D/2D contacts. a,** Perspective side view of a WSe$_2$ FET with degenerately *p*-doped WSe$_2$ (Nb$_{0.005}$W$_{0.995}$Se$_2$) contacts. **b,** Optical micrograph of the device. The channel region is encapsulated in hexagonal BN (h-BN) from the top and the bottom. **c,** Band profiles in the off- and the on-state of the FET with 2D/2D contacts. Holes are injected from a metal (M) into a degenerately *p*-doped WSe$_2$ contact (C) layer through a highly transparent M/C interface. Hole injection from the drain/source contact (C) layer across the 2D/2D interface into the undoped WSe$_2$ channel is modulated by the gate voltage $V_{bg}$. **d**, Modulation range of the source-drain current $I_{ds}$ by the back-gate voltage $V_{bg}$ yields an on/off ratio of up to >$10^9$ at room temperature in a WSe$_2$ FET with a 10.8 µm long and a 3.0 µm wide channel. **e,** Linearity of the $I_{ds}$-$V_{ds}$ characteristics indicating ohmic behavior for a wide range of back-gate voltages.

**Figure 2| Performance of a multilayer WSe$_2$ FET with degenerately *p*-doped WSe$_2$ (Nb$_{0.005}$W$_{0.995}$Se$_2$) 2D/2D contacts. a,** Schematic side view of a test structure for TLM measurements of the 2D/2D contact resistance. **b,** Optical micrograph of the corresponding WSe$_2$ TLM structure. The contours of the WSe$_2$ channel are marked by the dash-dotted lines, and those of the Nb$_{0.005}$W$_{0.995}$Se$_2$ contacts by dashed lines. Ti/Au electrodes for electrical connections are deposited on top of the Nb$_{0.005}$W$_{0.995}$Se$_2$ contacts. The thickness of the WSe$_2$ channel is ~ 7.0 nm. Scale bar, 1 µm. **c**, Total resistance *R* (multiplied by the channel width) versus channel length *L*. The intercept of the linear fit on the vertical axis yields the contact resistance 2$R_C$. **d,** Output characteristics of the shortest channel (*L*≈0.27 µm) in the TLM

structure. The maximum current exceeds 320 $\mu A/\mu m$ at $V_{ds}$ = - 1.5 V and $V_{bg}$ = -130 V.

**Figure 3| Observation of intrinsic channel properties in WSe$_2$ (a,c) and MoS$_2$ (b,d) devices with 2D/2D contacts. a,** Temperature-dependent two-terminal conductivity σ of WSe$_2$ as a function of the back-gate voltage $V_{bg}$ at $V_{ds}$ = -50 mV. The WSe$_2$ channel is ~ 3.5 nm thick, 14.8 µm long and a 4.7 µm wide. Inset: temperature dependence of σ at gate voltages ranging from -30 V to -80 V in steps of –5 V. A metal-insulator transition (MIT) is observed at ~$e^2/h$ as indicated by the dashed line. **b**, Temperature-dependent two-terminal conductivity σ of MoS$_2$ as a function of $V_{bg}$ at $V_{ds}$=-10 mV. The MoS$_2$ channel is ~ 6.8 nm thick, 13.0 µm long and a 2.7 µm wide. Inset: conductivity within the MIT region on an expanded scale. **c,d** Two-terminal field-effect hole mobilities $\mu_{FE}$ in WSe$_2$ (**c**) and MoS$_2$ (**d**) as a function of temperature. The maximum values observed in two-terminal measurements are $\mu_{FE}$≈2.0×10$^3$ cm$^2$V$^{-1}$s$^{-1}$ in WSe$_2$ and $\mu_{FE}$ ≈2.8×10$^3$ cm$^2$V$^{-1}$s$^{-1}$ in MoS$_2$ at cryogenic temperatures. The linearity of the output characteristics of WSe$_2$ and MoS$_2$ devices, shown in the insets of (**c,d**), indicates the absence of Schottky barriers in the contact region.

**Figure 4 | Characteristics and working principle of *p*-type field-effect transistors with 2D/2D hetero-contacts. a** Ohmic behavior observed in output characteristics of a MoSe$_2$ device with degenerately *p*-doped MoS$_2$ (Nb$_{0.005}$Mo$_{0.995}$S$_2$) drain/source contacts at 80 K, where the MoSe$_2$ channel is ~ 6.2 nm thick, 12.3 µm long and a 5.5 µm wide. **b,** Non-ohmic behavior displayed by MoSe$_2$ devices with degenerately *p*-doped WSe$_2$ (Nb$_{0.005}$W$_{0.995}$Se$_2$) drain/source contacts at 80K, where the MoSe$_2$ channel is ~ 5.0 nm thick, 6.3 µm long and a 2.1 µm wide. **c,** Two-terminal conductivity σ in the MoSe$_2$ device of **a** with *p*-doped MoS$_2$

contacts as a function of back-gate voltage $V_{bg}$ at different temperatures and $V_{ds}$ = -10 mV. Inset: two-terminal field-effect hole mobility as a function of temperature in the MoSe$_2$ device contacted by $p$-doped MoS$_2$. **d**, Conductivity σ in the MoSe$_2$ device of **b** with $p$-doped WSe$_2$ contacts as a function of $V_{bg}$ at different temperatures and $V_{ds}$ = -50 mV. **e,f** Changes in lateral band profiles induced by forming optimum 2D/2D heterojunctions in MoSe$_2$ contacted by $p$-doped MoS$_2$ (**e**) and $p$-doped WSe$_2$ (**f**).

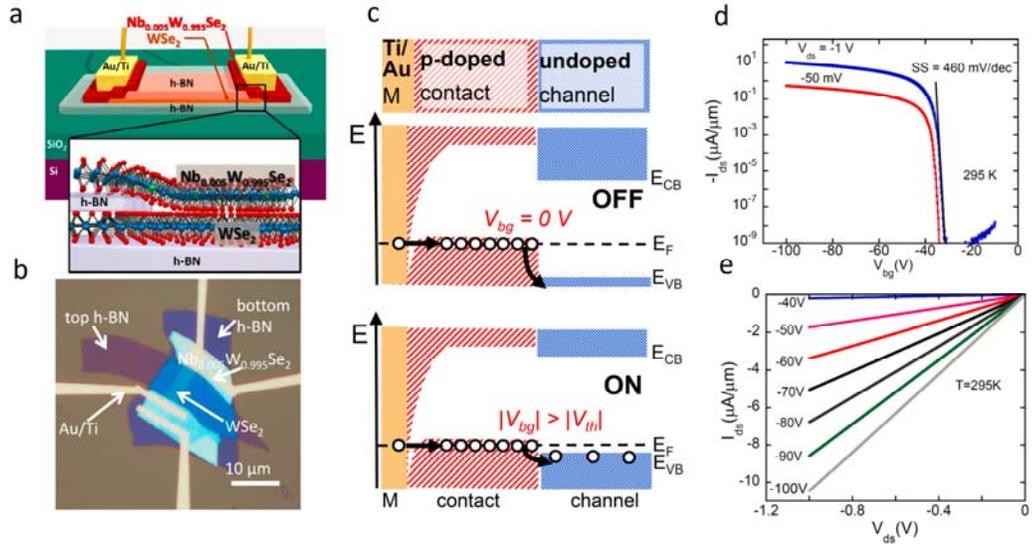

Figure 1

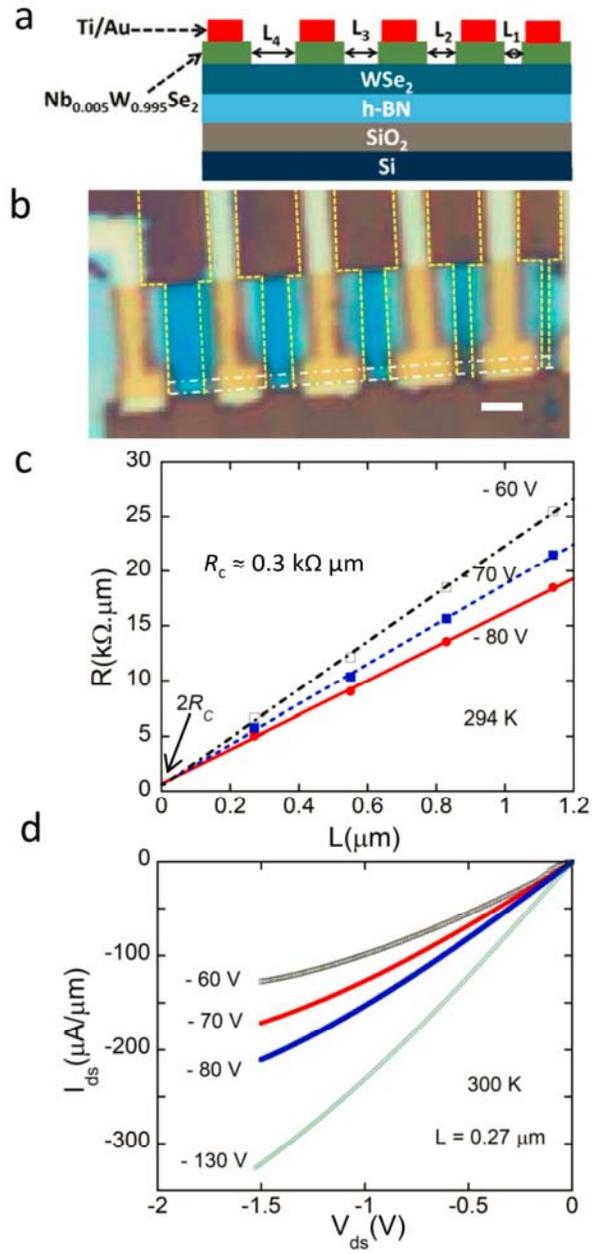

Figure 2

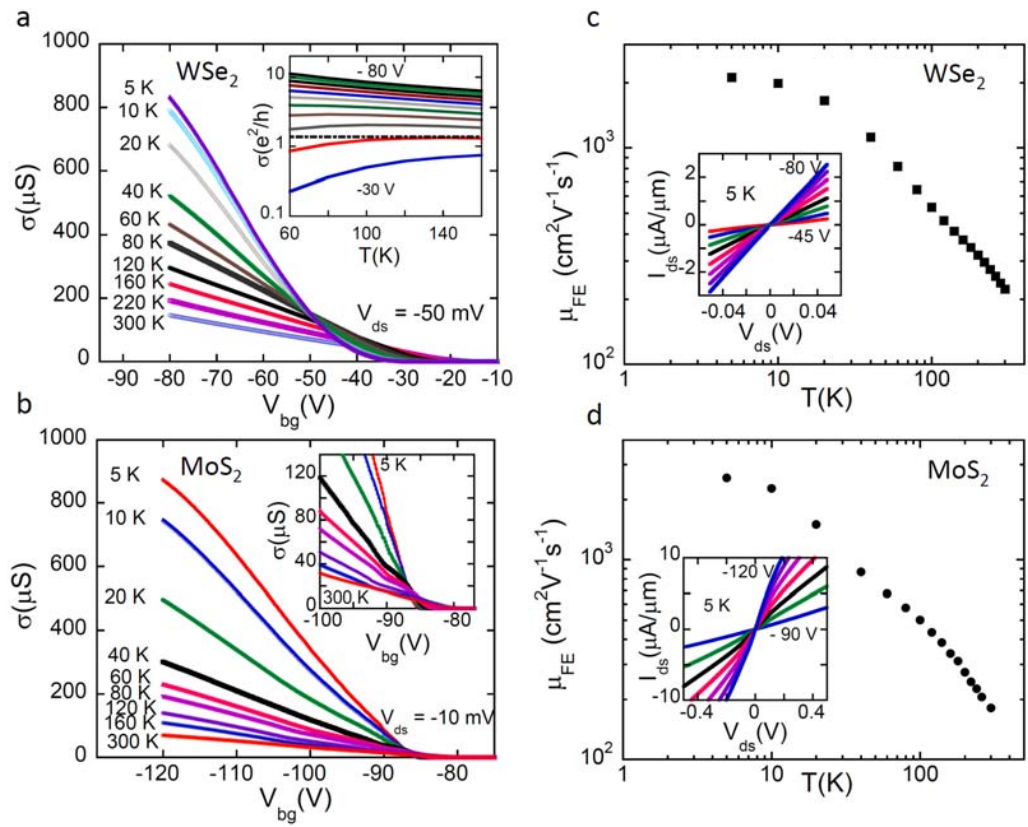

Figure 3

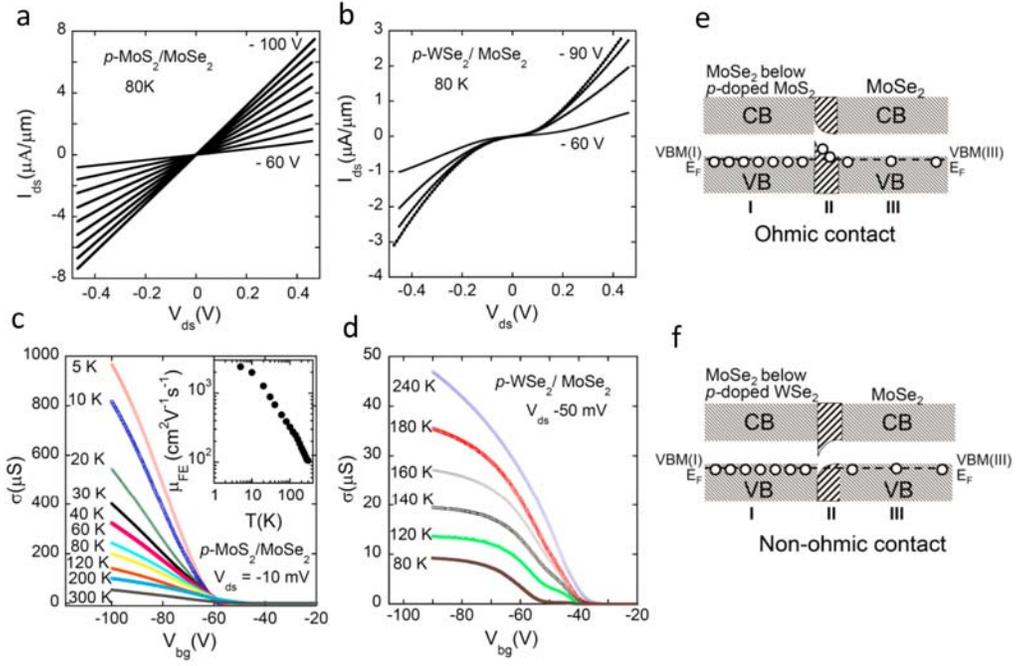

Figure 4

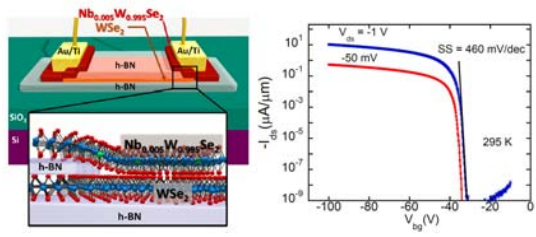

TOC